\documentclass{natureastroph}
\usepackage{geometry}                
\usepackage{graphicx}
\usepackage{amssymb}
\usepackage{epstopdf}
\title{Early assembly of the most massive galaxies}

\author{Chris A. Collins$^{1}$, John P. Stott$^{1}$, Matt Hilton$^{1, 2, 3}$, Scott T. Kay$^4$, S. Adam Stanford$^{5, 6}$, Michael Davidson$^7$, Mark Hosmer$^{8}$, Ben Hoyle$^9$, Andrew Liddle$^{8}$, Ed Lloyd-Davies$^{8}$, Robert G. Mann$^{7}$, Nicola Mehrtens$^{8}$, Christopher J. Miller$^{10}$, Robert C. Nichol$^{9}$, A. Kathy Romer$^{8}$, Martin Sahl\'{e}n$^{8}$, Pedro T. P. Viana$^{11, 12}$ \& Michael J. West$^{13}$}

\begin{document}

\maketitle

\begin{affiliations}
\item {Astrophysics Research Institute, Liverpool John Moores University, Twelve Quays House, Egerton Wharf, Birkenhead CH41 1LD.  
 \item Astrophysics and Cosmology Research Unit, School of Mathematical Sciences, University of KwaZulu-Natal, Westville Campus, Private Bag X54001, Durban 4000, South Africa. 
 \item South African Astronomical Observatory, PO Box 9, Observatory, 7935, Cape Town, South Africa.  
 \item Jodrell Bank Centre for Astrophysics, School of Physics and Astronomy, The University of Manchester, Manchester M13 9PL. 
  \item University of California, Davis, CA 95616, USA. 
  \item Institute of Geophysics and Planetary Physics, Lawrence Livermore National Laboratory, Livermore, CA 94551, USA. 
 \item SUPA, Institute of Astronomy, University of Edinburgh, Royal Observatory, Blackford Hill, Edinburgh, EH9 3HJ, UK. 
 \item Astronomy Centre, University of Sussex, Falmer, Brighton, BN1 9QH. 
 \item ICG, University of Portsmouth, Portsmouth PO1 2EG, UK. 
 \item Cerro-Tololo Inter-Amercian Observatory, National Optical Astronomy Observatory, 950 North Cherry Avenue, Tucson, AZ 85719, USA. 
 \item Departmento de Matem\'{a}tica Aplicada da Faculdade de Ci\^{e}ncias da Universiadade do Porto, Rua do Campo Alegre, 687, 4169-007 Porto, Portugal. 
  \item Centro de Astrofisica da Universidade do Porto, Rua das Estrelas, 4150-762 Porto, Portugal. 
  \item European Southern Observatory, Alonso de C\'{o}rdova 3107, Vitacura, Casilla 19001, Santiago 19, Chile.}
\end{affiliations}
\begin{abstract}
{\bf 
The current consensus is that galaxies begin as small density fluctuations in the early Universe and grow by in situ star formation and hierarchical merging \cite{sp05}. Stars begin to form relatively quickly in sub-galactic sized building blocks called haloes which are subsequently assembled into galaxies. However, exactly when this assembly takes place is a matter of some debate\cite{kam08, vd08}. Here we report that the stellar masses of brightest cluster galaxies, which are the most luminous objects emitting stellar light, some 9 billion years ago are not significantly different from their stellar masses today. Brightest cluster galaxies are almost fully assembled ${\bf4-5}$ Gyrs after the Big Bang, having grown to more than $\bf90\%$ of their final stellar mass by this time. Our data conflict with the most recent galaxy formation models \cite{lb07, b06} based on the largest simulations of dark matter halo development \cite{sp05}. These models predict protracted formation of brightest cluster galaxies over a Hubble time, with only  ${\bf 22\%}$ of the stellar mass assembled at the epoch probed by our sample. Our findings 
suggest a new picture in which brightest cluster galaxies experience an early period of rapid growth 
rather than prolonged hierarchical assembly. 
}
\end{abstract}

Brightest cluster galaxies (BCGs) are located at the centres of galaxy clusters. They constitute a separate population from bright elliptical galaxies \cite{vl08} and both their homogeneity and extreme luminosity have motivated their use as standard candles for cosmology \cite{san76,lp94, cm98}.
Our investigation focuses on BCGs in the most distant X-ray emitting galaxy clusters at redshifts $z=1.2-1.5$, where $(1+z)$ is the expansion factor of the Universe relative to the present. It has been shown that X-ray cluster selection is currently
the optimum strategy for an unbiased investigation of BCG evolution \cite{bcm00}. 
 Properties of our BCGs and their host clusters are listed in Table~1. All five clusters were discovered serendipitously in X-rays and they are the most distant clusters discovered in their respective X-ray surveys \cite{brem06, stan06, ros99, demarco07, mullis05}. The cluster J2215 was discovered as part of the XMM Cluster Survey  (XCS \cite{rom01, sh08}) and has the highest redshift of any spectroscopically confirmed cluster \cite{stan06, hil07}.  

The stellar mass of a BCG depends upon the hierarchical build up of its host dark matter halo and its stellar evolution history, along with the baryonic physics of the galaxy.
We base our study of BCGs on photometry in the infrared wavebands $J\, (1.26\, \mu \rm{m})$ and $K_s \,(2.14\, \mu \rm{m})$. Infrared imaging is essential at these large redshifts to compensate for the redshifting of the early-type galaxy spectra. Also, these wavebands are less sensitive than optical light to the presence of young stars and are a more accurate tracer of the underlying old stellar population and, hence, of the stellar mass of the systems. Fig.~1 shows an infrared image of the cluster J2235 from our sample (see also Supplementary Fig.~1).  
 

We start by examining the ages of the stars themselves in these galaxies using the run of $J-K_s$ colour evolution with redshift as shown in Fig.~2.  
For BCGs at the redshift of our sample the $J-K_s$ colour predictions for the models separate clearly. For the comparison sample at lower redshift we use X-ray selected clusters \cite{stott08} which are well matched in mass to our own cluster sample. There is a remarkable agreement between the data and the hybrid model (see Fig.~2 legend), with all five BCGs lying within $0.05~$mags of their predicted colour, indicating a consistent epoch of formation for the majority of the constituent stars in all systems between redshifts $z_f=3-5$, some 2-3~Gyr after the Big Bang.

Turning our attention to the mass assembly of BCGs implied by our data, in Fig.~3 (see also Supplementary Table~1 and Supplementary Fig.~2) we show the estimates of stellar mass for our distant BCGs normalised to the average mass of the comparison sample at $z\leq 0.04$, which is $8.99\, (\pm 0.82) \times 10^{11}\, $M$_{\odot}$ (s.e.m.). Using a Tukey's biweight location estimator for robustness, for our five objects located at $z=1.22-1.46$ we find an average stellar mass of $8.86\, (\pm 1.73) \times 10^{11} \,$M$_\odot$ (s.e.m.). The ratio of these estimates is $0.99\pm0.21$ (s.e.m.), indicating that on average the masses of the high redshift BCGs are consistent with local counterparts.

To compare with theory we use the haloes from the Millennium\,Simulation\cite{sp05}\\ (http://www.mpa-garching.mpg.de/millennium) matched to the total mass of our clusters, estimated from their X-ray luminosity (see Supplementary Information). The mass range of our five clusters (Table 1) has excellent overlap with the combined $z=1.08$ and $z=1.5$ halo samples  \cite{lb07} (Supplementary Fig.~3). The predicted hierarchical mass build up of BCGs in these 250 haloes is also shown in Fig.~3. The corresponding mass of the simulated BCGs has grown to an average of only $1.92\,(\pm 0.38) \times 10^{11} \,$M$_{\odot}$ (s.d.) by this time, some $22\%$ of the observed value. The data are inconsistent with the prediction at  the level of $4 \sigma$ (one-tailed $P=0.008$, degrees of freedom $\rm{d.f.}=4$; based on a Student's t distribution appropriate for small samples).


To check the stability of the BCG assembly predictions we selected massive haloes from the independent Durham semi-analytic model \cite{b06} which also uses the Millennium Simulation\cite{sp05} but incorporates a different treatment of the baryon physics close to active galactic nuclei, partly in order to better reproduce the abundance of massive elliptical galaxies at high redshift.  Using the same  selection limits we find that the BCG mass fractions compared to the present day are  $0.22^{+0.18}_{-0.09}$ at $z=1$ and $0.17^{+0.12}_{-0.07}$ at $z=1.5$, indicating good agreement between the two semi-analytical models.

It is well known that the estimates of stellar mass from photometry even for early-type galaxies such as BCGs depend on the underlying stellar evolution model used. To investigate this sensitivity we have applied three independent stellar population synthesis codes to early-type galaxies at the mean redshift of our sample ($z=1.3$) using a range of model parameters (see Supplementary Table 1).   
These results show that the $K_s$ band stellar mass estimates remain significantly discrepant from the semi-analytic predictions (one-tailed $P \leq 0.02, \rm{d.f.}=4$) for the vast majority of parameters considered across the three models, reaching  a value for one-tailed $P\, \rm{of} \geq 0.05$ in one of the three only if the stellar formation epoch $z_f$ is less than 2.5 together with a stellar metallicity less than the solar value. This situation is incompatible with observations of BCGs and massive early-type galaxies in general (see Supplementary Information).  
We conclude that there remains a significant discrepancy between the recent semi-analytic models of galaxy formation coupled to the largest N-body simulations and the stellar masses of BCGs at the centres of the most massive clusters.

In comparison to recent studies \cite{whey08}, this work significantly extends the redshift baseline over which BCG evolution has been investigated to $z=1.5$, equivalent to a look-back time $\simeq65\%$ the age of the Universe. Although the first glimpse of the $z>1$  BCG population reveals galaxies with a range of stellar masses, there is on average considerably less stellar mass evolution than expected, with the bulk ($\geq90\%$) of the stellar mass already in place by $z\simeq1.5$, corresponding to only 
about 4-5~Gyrs after the Big Bang; the
current models predict a considerably longer timescale of about 11~Gyr for the same growth, reaching $90\%$ at $z\simeq0.2$. 


Despite this there is evidence that merging is still underway in our high redshift sample. The BCG in J0849 at $z=1.26$ has a nearby companion (projected separation of about 6\, kpc) with which it is likely to undergo dissipationless merging in the future \cite{yamada2002}. Of the other clusters in our sample, the BCG and its neighbour (projected separation of about 15\,kpc) in J1252 are also possible merger candidates. 
Assuming that mergers take place in both these cases, the fraction of BCG stellar mass already assembled (based on the $K_s$ fluxes of the main components) is $\simeq84\%$ and $\simeq 60\%$ for J0849 and J1252 respectively, supporting the contention that 
most of the growth has actually already taken place in these two BCGs. 

The timescale for the mass assemblage is similar to the age of the component stars ($2-3$~Gyrs), a situation that appears to resembles classical monolithic collapse \cite{ebs62, lar74} rather than hierarchical formation. To form a galaxy of stellar mass $10^{12}\,$ M$_{\odot}$ over 4 Gyrs requires a mass deposition rate of about 250 \,M$_{\odot}$ yr$^{-1}$ and an efficient mechanism to feed the gas into the inner regions of the halo where it can form stars. Unfortunately the merging process becomes inefficient for massive galaxies because merger induced shocks lead to heating as opposed to radiative cooling of the gas \cite{binney04}. One recent suggestion \cite{dekel09} is that the early assembly of massive galaxies at $z\geq 2$ is driven by narrow streams of dense cold gas which penetrate the shock-heated region greatly increasing the efficiency of the gas deposition and associated star formation. Thus, the fraction of time that young BCGs spend undergoing a major merger event could be $\leq 10\%$, with the stellar mass assembly dominated by this `stream-fed' process \cite{dekel09}. Alternatively, a deficiency may lie in the semi-analytic treatment of the physical processes in the densest environments during early hierarchical assembly; this contention is supported by the fact that current predictions are moderately consistent with observations of the evolution of luminous red galaxies \cite{wake06, alm08}, whereas our results, which focus on the most massive subset of this population, the BCGs, differ much more from the model predictions.  

In a wider context the hierarchical simulations and their semi-analytic prescriptions have arguably provided an excellent way of generating mock catalogues of galaxies to compare with real data, but our results show that they do not account for the assemblage history of all galaxies. Larger simulations may provide a better statistical probe of both the merging history of the largest haloes and cluster-mass trends. If BCGs collapsed and formed at high redshift in a single burst of intense star formation then they may well be dusty and in sufficient numbers to be detectable with the coming generation of submillimetre surveys, which will cover areas large enough to detect objects as rare as BCGs. The ongoing XCS survey will find many more high redshift clusters and we anticipate that our results will stimulate independent  studies of BCGs as new clusters are found in the redshift `desert' beyond $z=1.5$ from infrared and X-ray based surveys such as eRosita.

\newpage




\begin{addendum}

\item[Supplementary Information] is linked to the online version of the paper at www.nature.com/nature.

 \item This work is based in part on data collected at the Subaru Telescope, which is operated by the National Astronomical Observatory of Japan and the XMM-Newton, an ESA science mission funded by contributions from ESA member states and from NASA. We acknowledge financial support from Liverpool John Moores University and the STFC. M.H. acknowledges support from the South African National Research Foundation. IRAF is distributed by the National Optical Astronomy Observatories, which are operated by the Association of Universities for Research in Astronomy, Inc., under cooperative agreement with the National Science Foundation. We thank  Gabriella De Lucia for making simulation results available to us in tabular form, Ichi Tanaka for developing the MCSRED package used to reduce the MOIRCS data, Maurizio Salaris for discussions on stellar population synthesis models and Ben Maughan for discussions on cluster masses. 

\item[Author Contributions] C.A.C. provided the scientific leadership, helped design the experiment, wrote the paper and led the interpretation. J.P.S. performed the photometry and data analysis and made major contributions to the interpretation. M.H. wrote the Subaru proposal, carried out the data reduction and photometric calibration, contributed to the analysis and interpretation and provided detailed comments on the manuscript. S.T.K. independently checked the cluster mass calculations. S.A.S. provided useful discussions on the data and comments on the manuscript. The remaining authors make up the team of the wider XCS project which led to the discovery of J2215. R.G.M, R.C.N, and A.K.R. made useful comments on the text.
 
  \item[Competing Interests] The authors declare that they have no
competing financial interests.
 \item[Correspondence] Correspondence and requests for materials
should be addressed to\\ C.A.C. (email: cac@astro.livjm.ac.uk).
\end{addendum}

\newpage

\noindent
\newpage
\begin{figure}
\includegraphics[bb=-60 0 200 225]{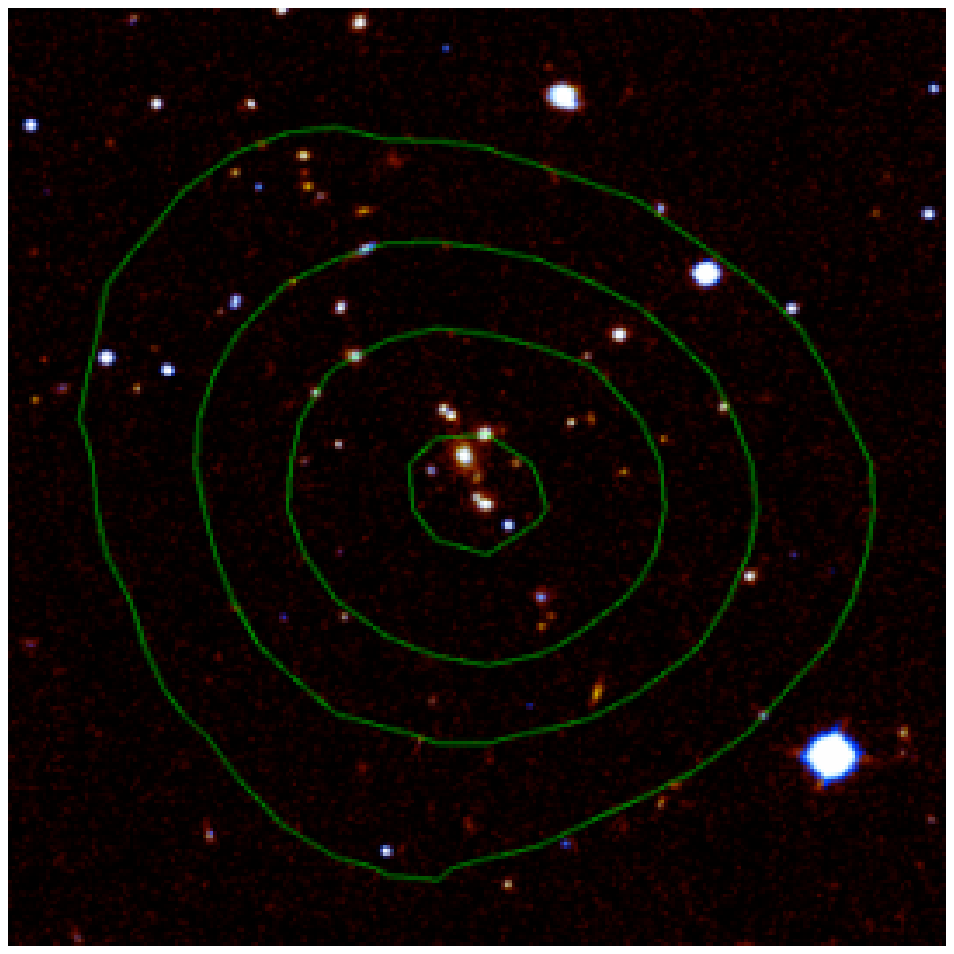}\\ 
\noindent
{\small{\bf Figure 1. Infrared image of the cluster J2235.} An infrared image of the cluster J2235 at a redshift $z=1.39$. Data were taken using the 8.2-m Subaru telescope. The image is combined from separate $J$ and $K_s$ exposures and shows the $1.5' \times 1.5'$ region surrounding the cluster centre. At this redshift $1.5'$ corresponds approximately to 0.75~Mpc. The green overlaid contours show the smoothed X-ray emission taken from the XMM-Newton XCS pipeline, smoothed with a Gaussian kernel. The X-ray peak coincides with the cluster centre and the position of the BCG. For a full description of the observations and data reduction see Supplementary Information.}
\end{figure}

\newpage
\begin{figure}
\includegraphics[bb=100 40 250 250]{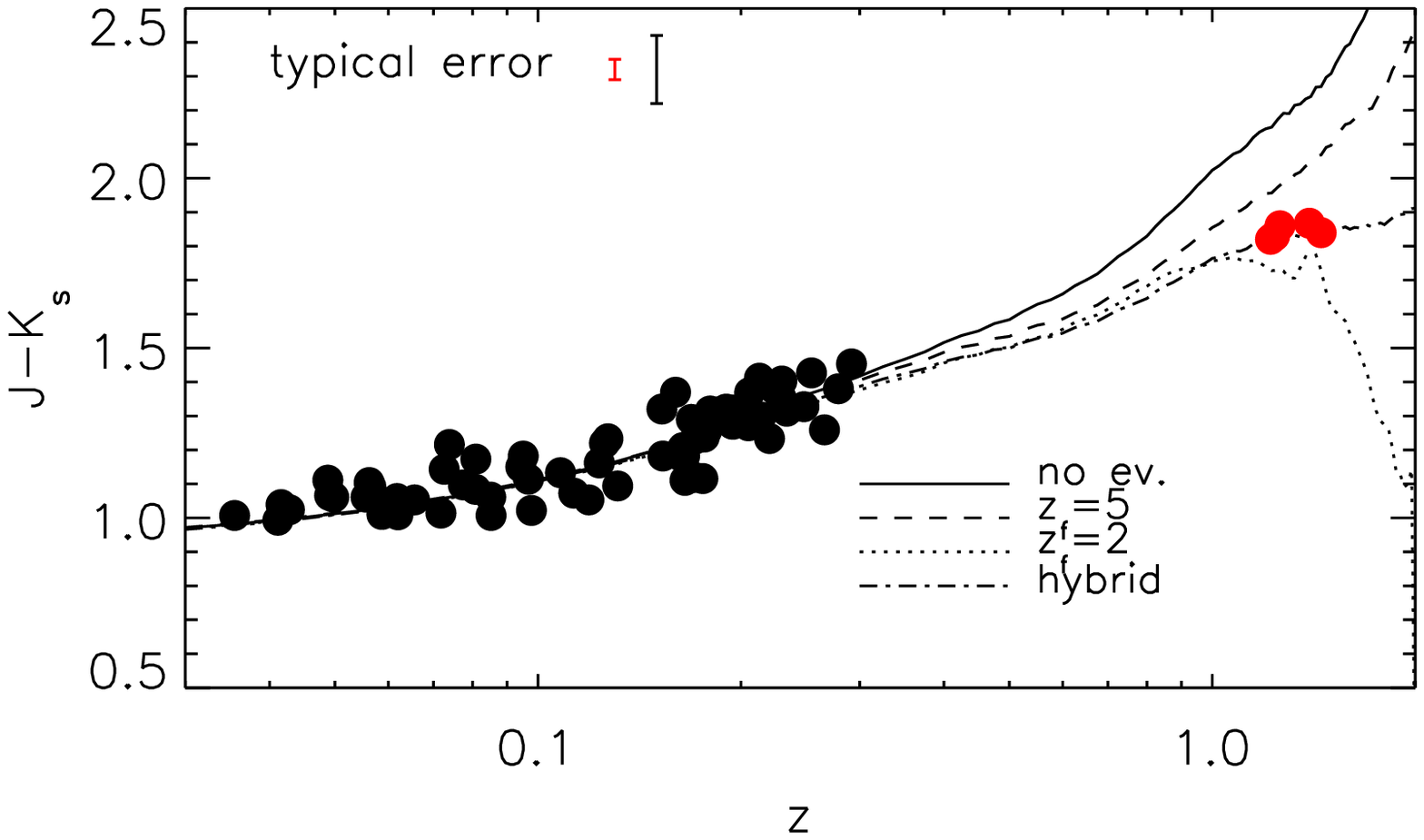} 

\noindent
{\small{\bf Figure 2. The stellar evolution of BCGs with redshift.} The $J-K_s$ colour evolution for our five high redshift BCGs (red) and 72 BCGs from the comparison sample \cite{stott08} (black) which have host cluster masses in the same range as our high redshift clusters and have available $J$ and $K_s$ photometry.  The errors (s.d.) reported for the comparison sample \cite{stott08} and our data are $\simeq 0.1\,$ mag and $\simeq 0.02\,$ mag respectively and are shown in the figure. This plot includes simple stellar population models \cite{bc03} incorporating: no stellar evolution (solid); passive evolution with formation epoch $z_f=5$ (dashed); passive evolution with formation epoch $z_f=2$ (dotted);  a hybrid model with an exponentially decaying star formation rate in which $50\%$ of the BCG stellar content is formed by $z_f=5$ and $80\%$ by $z_f=3$ (dot-dashed), which is appropriate to the star formation history predicted by the semi-analytic model \cite{lb07}. The $z_f=2$ and $z_f=5$ stellar models are calculated  assuming solar metalicity and  a Salpeter initial mass function (IMF) \cite{bc03}, while the hybrid model was calculated with a Chabrier IMF \cite{ch03}. The implied epoch of formation $z_f=3-5$ ($2-3$ Gyrs after the Big Bang) agrees well with other estimates of stellar ages determined for BCGs and early-type galaxies in clusters (see Supplementary Information). Throughout our analysis we assume a concordance cosmology of $\Omega_m=0.3, \Omega_{\Lambda}=0.7,$ and $H_0=70$ km s$^{-1}$ Mpc$^{-1}$, where $\Omega_{\Lambda}$ is the energy density associated with a cosmological constant. See the Supplementary Information for details of data reduction.} 
\end{figure}


\newpage
\begin{figure}
\includegraphics[bb=85 10 150 240]{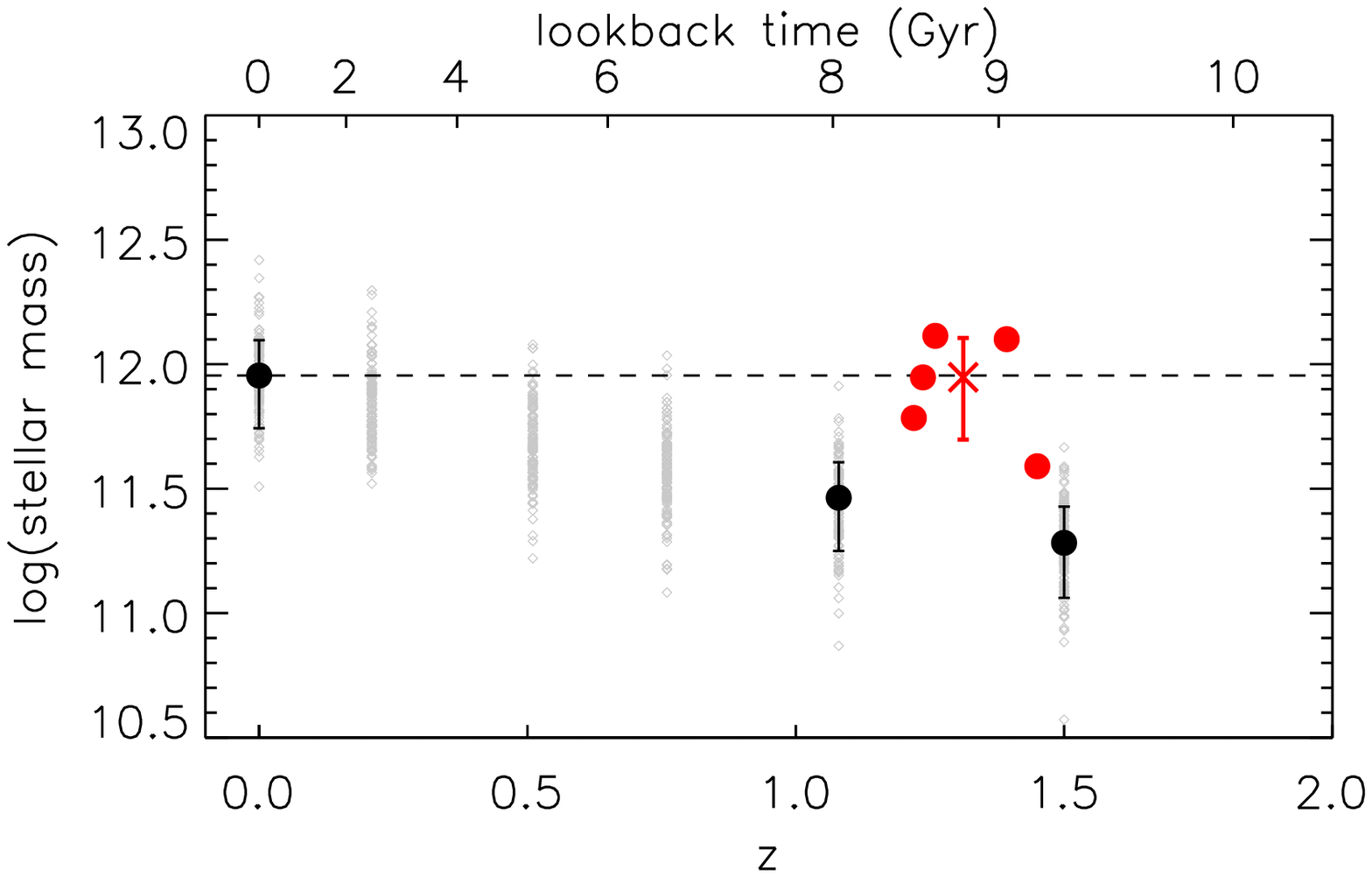}\\ 
\noindent
\small{\bf Figure 3. The mass evolution of BCGs with redshift.} The BCG mass estimates of our sample normalised to local galaxies at $z=0.04$. The red cross is the estimated biweight  location ($8.86 \times 10^{11}\,${M$_{\odot}$) and scale ($3.87 \times 10^{11}\,$M$_{\odot}$) of the sample.
 We calibrate the stellar masses by comparing the rest-frame absolute $K_s$ magnitudes with the predicted magnitudes and corresponding stellar masses from the semi-analytic models \cite{lb07}. This involves correcting the observed $K_s$ for: cosmological dimming; sampling different spectral regions of the galaxies resulting from the redshift ($k$-correction); and stellar evolution. The last two corrections are carried out using synthesized stellar spectra for early-type galaxies (appropriate to BCGs) from the hybrid stellar population model shown in Fig.~2. The $k$-correction is well understood over the wavelength range appropriate to our sample ($0.9-2.2\, \mu \rm{m}$) introducing an uncertainty of about 10\% in the rest-frame absolute $K_s$ magnitude estimates. The biweight scale provides a realistic estimate of the intrinsic error (s.d.) in the average mass using the hybrid model, however the total uncertainty in the inferred BCG mass is larger as it depends on the stellar evolution model used (see Supplementary Information). The grey diamonds show the individual BCG mass predictions \cite{lb07} in 125 simulated clusters at each of six redshifts (0.0, 0.2, 0.5, 0.75, 1.08, 1.5) above corresponding selection masses (4.7, 3.5, 2.8, 2.4, 1.5, 1.0) in units $10^{14}\,$M$_{\odot}$. The black filled circles show the average value at each redshift (all errors are s.d.).  The predictions are based on semi-analytic models of galaxy evolution. These use large N-body simulations such as the Millennium Simulation \cite{sp05}, which models the development of  $2160^3$ cold dark matter particles within a box over 2 billion light years on each side. The semi-analytic techniques use the merger trees from the simulations and graft on analytical approximations to account for the complicated physics of the baryons in a range of ongoing processes associated with galaxy formation, such as: cooling, star formation, supernova outbursts and the growth of black holes in active galactic nuclei.}
\end{figure}


\begin{table*}\scriptsize
\begin{center}
{\bf\small Table 1: The properties of the host clusters and their BCGs} \label{tbl-1}
\vspace{14pt}
\begin{tabular}{lcccccr}
\hline
Cluster Name &  ${\rm Redshift}\, $ & X-ray luminosity & Cluster Mass &BCG $K_s$ (total) &$J-K$& Stellar Mass\\
&      &    ($10^{44}$~erg s$^{-1}$)     &($10^{14}\,$M$_{\odot}$) & & &($10^{12}\,$M$_{\odot}$) \\
\hline
XLSS J022303.0-043622 (J0223) & 1.22 & $1.1^{+0.1}_{-0.1}$&$1.0\pm0.4$ &$17.72\pm0.01$& $1.82\pm0.01$&$0.61\pm0.08$\\
XMMU J2235.3-2557 (J2235) & 1.39 & $11.4^{+0.7}_{-0.7}$& $3.1\pm0.7$&$17.34\pm 0.01$&$1.87\pm0.02$ & $1.26\pm0.14$\\
XMMXCS J2215.9-1738 (J2215) &  1.46 & $4.4^{+0.8}_{-0.6}$&$1.8\pm0.4$ &$18.72\pm 0.01$&$1.83\pm0.02$ &$0.39\pm0.05$\\
RX J0848.9 + 4452 (J0849) & 1.26&$3.3^{+0.9}_{-0.5}$&$1.8\pm0.4$ &$17.00 \pm 0.02$&$1.86\pm0.03$ & $1.30\pm0.15$  \\
RDCS J1252.9 -2927 (J1252) & 1.24&$6.6^{+1.1}_{-1.1}$&$2.6\pm0.6$ &$17.36 \pm 0.03$&$1.83\pm0.01$& $0.89\pm0.11$\\
\hline
\end{tabular}
\vspace*{15pt}
\end{center}
{\small The cluster X-ray luminosities are bolometric estimates taken from the literature and the cluster masses are $M_{200}$ values (Supplementary Information). The errors on the cluster masses are based on the X-ray luminosity errors and the intrinsic uncertainty in the scaling relations. The $J$ and $K_s$ observations of  J0223, J2235 and J2215 (Supplementary Fig.~1) were taken with the 8.2-m Subaru telescope and reach a $5\sigma$ (s.d.) limiting magnitude $J\simeq23.7$ and $K_s\simeq 22.8$ (23.1 in the case of J0223). The photometry for our data was calibrated using standard stars taken on the night in the Vega system. For comparison with previous observations we find that our J0223 BCG total $K_{s}$-band magnitude ($K_s=$17.72$\pm$0.01) is in excellent agreement with the total magnitude from the literature \cite{brem06} ($K_s=$17.76$\pm$0.04, assuming a $K_{s}$-band conversion from AB to Vega system of -1.86).  The photometry for J1252 and J0849 were sourced from the literature \cite{demarco07, yamada2002} and for these galaxies the total $K_s$ magnitudes and $J-K_s$ colours were measured in similar aperture sizes. All data have been analysed in an identical manner for direct comparison (see Supplementary Information). The errors on the stellar masses include all photometric errors and the uncertainty in the calibration with the semi-analytic model \cite{lb07}. All errors are $1 \sigma$ (s.d.). For each cluster we identified the brightest galaxy from the $K_{s}$-band magnitudes of all galaxies within 500~kpc of the cluster X-ray centroid because for approximately $95\%$ of clusters the BCG lies within this radius \cite{lm04}. All identified BCGs have optical spectra confirming their cluster membership  \cite{brem06, hil07, ros99, demarco07, mullis05}.
}
\end{table*}
\newpage


\pagebreak

{\noindent\bf\Large Early assembly of the most massive galaxies}\\
{\noindent\bf\Large Supplementary Information}

\begin{abstract}
{ \vspace{20pt} 
We provide further details of the observations and data reduction; carry out a comparison of the cluster masses used in the Millennium Simulation with those of our high redshift sample; and we also discuss the BCG mass determinations in greater detail and how these are affected by uncertainties in the stellar formation history of the galaxies.}
\end{abstract}
\vspace{20pt}



\noindent{\bf Observations and Data Reduction}

The MOIRCS  instrument on the Subaru Telescope provides imaging and low-resolution spectroscopy over a total field-of-view of $4'\times7'$ with a pixel scale of $0.117''$ per pixel.  
This is achieved by dividing the Cassegrain focal plane and then re-focussing the light through identical optics onto two HAWAII-2 $2048 \times 2048$ CCDs, each covering $4'\times3.5'$. Observations were taken in photometric conditions in $0.5''$ seeing on the nights of August 8th and 9th 2007, with the clusters centred on Detector 2. A circular 11-point dither pattern of radius $25''$ was used for both bands to ensure good sky subtraction. Integration times were 25 mins at $J$ and 21 mins at $K$$_{s}$ (37 mins for J0223). These exposures reach a $5 \sigma$ (s.d.)  limiting magnitude of $J \simeq 23.7$ and $K_{ s} \simeq 22.8$ (23.1 in the case of J0223). Supplementary Fig.~1 shows combined $J$ and $K$$_{ s}$ images of J0223, J2235, and J2215.

The data were reduced using the external IRAF package MCSRED. The data were flat fielded, sky subtracted, corrected for distortion caused by the camera optical design, and registered to a common pixel coordinate system. The final reduced images on which we performed the photometry were made by taking the 3$\sigma$ (s.d.) clipped mean of the  dither frames. The BCG photometry was extracted in an identical manner to the comparison sample using SExtractor (version 2.5)  `Best' magnitude, which is found to be within 0.1 magnitude of the total. Measuring total magnitude is a significant improvement over the fixed aperture photometry of previous studies \cite{cm98, whey08} and constitutes the optimum comparison with semi-analytic models, which provide no information on the spatial distribution of light from merging haloes.  Furthermore in the densely populated regions such as cluster cores the `Best' magnitude initially reverts to the isophotal magnitude (MAG ISO), enabling light from close neighbours to be excluded, and then incorporates an empirical aperture correction. To calculate the colours of the BCGs we run SExtractor in dual mode so that the $K$$_{s}$ band detections extract the $J$ band catalogue with identical positions and apertures which ensures accurate colour determination.


\pagebreak   
 \noindent{\bf Cluster Masses}
 
  Various authors have identified  a weak correlation between BCG mass and their host cluster mass \cite{e91,cm98, br08, whey08} which does not change significantly with redshift out to $z\simeq0.8$. If cluster masses have grown by a factor 2 or 3 since $z\simeq1$, as some authors suggest \cite{li07}, then their BCGs would be expected to grow by $20-30\%$ over the same time interval. Therefore in order to compare BCG masses in a meaningful way it is necessary that our cluster sample is well matched to the masses of simulated clusters in the Millennium Simulation with which we are comparing. The clusters of interest from the simulation are the 125 most massive systems at the two redshifts $z=1.08$ and $z=1.5$,  selected for comparison with observations \cite{lb07}. Halo masses $M_{200}$ are measured at a radius ($R_{200}$) inside which the average mass density is 200 times the critical density of the Universe. 
   
In order to compare our data with simulations we use the bolometric X-ray luminosity of each system ($L_{\rm{X}}$) \cite{brem06, stan06, stan01, rosati2004} as this can be used to determine cluster mass from power-law scaling relations \cite{kaiser1986, randb2002}. For J2235 we use the published X-ray luminosity \cite{mullis05} in the energy range $0.5-2.0$ keV and convert to bolometric luminosity assuming a temperature of 6 keV. The mass-observable scaling relations are reasonably well calibrated at low redshift but are far less well measured beyond $z\simeq0.5$, where evolution becomes important. However, it has been shown \cite{maughan2007, maughan2008} that self-similar evolution provides a reasonable description of the dynamical state of clusters and furthermore that simple luminosity is a reliable mass proxy, with an intrinsic scatter $\sigma_M=21\%$ (s.d.) out to $z \simeq 1$. We adopt the self-similar scaling relation between $L_{\rm{X}}$ and $M_{500}$ (where $M_{500}$ is defined in an analogous way to $M_{200}$) calculated for 115 clusters in the range $0.1 < z < 1.3$ (ref.~38) and given by
$$  L_{\rm{X}}=C E(z)^{\alpha} \left ( \frac{M_{500}}{4\times10^{14}} \right ) ^{\beta} \rm{erg\,s^{-1}},$$

where $C=5.6\pm 0.3\times10^{44}$ erg s$^{-1}$, $\alpha=7/3$, $\beta=1.96\pm0.10$ and $E(z)$ describes the evolution of the Hubble parameter;
$$E(z)=[\Omega_m(1+z)^3 + (1-\Omega_m - \Omega_{\Lambda})(1+z)^2 + \Omega_{\Lambda}]^{1/2}.$$ 


 Total X-ray luminosities for our clusters were measured using apertures of at least 50 arcsec radius, corresponding to about 500 kpc at $z=1$, which is close to $R_{500}$ for massive clusters. From the Millennium Gas simulation \cite{hartley2008} the ratio of $L_{\rm {X}}$ at $R_{200}$ and $R_{500}$ is only 1.03, which adds weight to the assumption that $L_{\rm {X}}$ measured at $R_{500}$ captures nearly all the X-ray emission. Finally, we calculate the $M_{200}$ values using the conversion from $M_{500}$ (ref. 41). In the absence of cool core clusters at high redshifts \cite{maughan2007} there is no need to exclude the core emission from these estimates. The $L_{\rm{X}}$ values and final $M_{200}$ mass estimates for our clusters are shown in Table~1 of the main text. 
 
 To demonstrate stability we note that using cluster temperature, where available, as a proxy for mass \cite{maughan2007}, gives virtually identical results. Independent mass measurements are available for some clusters which provide an important check on the self-similar scaling assumption. From a dynamical analysis of J2215 (ref.~18), the cluster velocity dispersion $\sigma_v=570\pm 190$ km s$^{-1}$ and  $R_{200}=0.63\pm0.15$ Mpc. Adopting the relationship \cite{carlberg1997}
 $$M_{200} = \frac{3}{G} \sigma_v^2 R_{200}$$
 
gives $M_{200} = 1.4^{+1.1}_{-0.8} \times10^{14}\,M_{\odot}$, which compares very well with the value in Table~1 of $1.8\,(\pm0.4) \times10^{14}\,M_{\odot}$.  From a weak-lensing analysis of the Lynx-East cluster J0849 \cite{jee2006}, $M_{200}=2.0\,(\pm 0.6)\times10^{14}\,M_{\odot}$, compared to our value of $1.8\,(\pm 0.4) \times10^{14}\,M_{\odot}$. Using the X-ray surface brightness of J1252 measured with Chandra and XMM-Newton \cite{rosati2004} gives a mass $M_{500} = 1.9\, (\pm 0.3) \times10^{14}\,M_{\odot}$, which compares favourably with our estimate measured at $R_{500}$ of $1.9\,(\pm0.6) \times10^{14}\,M_{\odot}$, corresponding to $M_{200}=2.6\,(\pm0.6) \times10^{14}\,M_{\odot}$. This mass also agrees with the estimate within $R_{500}$ from the weak-lensing analysis of the cluster \cite{lom2005}. Finally, an extensive optical spectroscopic survey of J1252 (ref.~14) reveals a mass inside $R_{500}$ of $1.6-2.3\times10^{14}\,M_{\odot}$, again consistent with our estimated value. 
We therefore conclude that our mass estimates are reasonably reliable and consistent between independent methods despite the high redshift of the sample.
 
Crucially these cluster masses are comparable to the massive haloes seen in the Millennium Simulation. A histogram of halo and cluster masses is shown in Supplementary Fig.~3. The simulated cluster samples at $z=1.08$ and $z=1.5$ have lower mass limits  at these redshifts of $1.5\times10^{14}\,M_\odot$ and $1.0\times10^{14}\,M_\odot$ respectively. The average mass of the combined high redshift sample of 250 haloes is $2.3\,(\pm1.1) \times 10^{14}\,M_{\odot}$ (s.d.), compared to the average mass for our sample of $2.1\,(\pm0.8) \times 10^{14}\,M_{\odot}$ (s.d.).
Although the predicted mass function is relatively steep and the halo numbers fall rapidly with increasing mass \cite{sp05}, there are still 32 clusters at $z=1.08$ and a further 6 at $z=1.5$ with a mass larger than our most massive cluster J2235.  

\noindent {\bf Stability of BCG Mass Estimates}

Most studies of the $K$ band Hubble diagram for BCGs have assumed that the near-IR light is a direct proxy for stellar mass \cite{cm98, abk98, sb02, nel02, whey08}, sometimes using colours to confirm the presence of an old stellar population. However, estimates of the stellar masses of galaxies are known to depend on the underlying stellar evolution model used, leading to degeneracy and systematic uncertainties. In particular some stellar evolution models \cite{mar2005, mar2006} are known to produce younger ages and smaller stellar masses compared to others. In order to investigate these systematic effects in our BCG mass estimates, to provide plausible errors and to identify the major assumptions on which our results depend, we study the effect of using a representative set of different stellar models covering an appropriate range of values for the important parameters. 

We concentrate on three simple-stellar population (SSP) synthesis models: Bruzual and Charlot \cite{bc03} (hereafter BC), which is based on the Padova stellar models and Geneva spectral libraries; Maraston\cite{mar2005} (hereafter MAR) and BaSTI \cite{piet2004, percival2009}. The last two stellar population models are based on independent spectral libraries and stellar tracks and generally incorporate a larger range of parameters. However, the significant addition in the MAR and BaSTI models, compared to BC, is that they both implement improved modeling of the thermally pulsating phase of extremely bright asymptotic giant branch (AGB) stars.
It is important for our purposes to include a model with a realistic prescription for AGB evolution because their contribution to the near-infrared light can be significant, even though the stars are relatively few in number, giving rise to significantly smaller mass-to-light $(M/L)$ ratios and therefore smaller stellar masses for a given $K_{s}$ luminosity. This AGB evolutionary phase is difficult to model in detail and can be a source of discrepancy between different population synthesis models, hence the inclusion of two independent predictions here. 

We consider the SSP models with a range of metallicities $Z=0.4\, Z_\odot$--$2.5\, Z_\odot$, where $Z$ is the total mass fraction of elements heavier than helium measured in solar units, and four formation redshifts $z_f=2,2.5,3,5$. For BC we use a Chabrier \cite{ch03} stellar initial mass function (IMF), while for those of MAR and BaSTI we use a Kroupa \cite{kroup2001} IMF. These IMFs are similar and both account for the flatter slope of low mass ($\leq1\,M_\odot$) stars as observed in star counts.  
The major difference between Chabrier and Kroupa is simply the parametric fit to the low mass slope of the IMF. 

We extract the appropriate $M/L$ ratios from the SSP codes for the above ages and metallicities, which gives us the variation in stellar mass for a given observed flux in the $K_{s}$ band and therefore a realistic handle on the uncertainty of the BCG masses. We normalise the $M/L$ ratios to corresponding values \cite{lb07} at $z=0$ and then compare the $M/L$ estimates using the appropriate age for the mean redshift of our sample ($z=1.3$). In Supplementary Fig.~4 we plot the BCG stellar mass at $z=1.3$ against metallicity and formation redshift by representing this quantity as a surface for each of the different stellar population codes and in Supplementary Table~1 we list each individual stellar mass. The range in stellar mass predicted by these models is about 0.3 dex. In general we find that the older populations have a higher mass-to-light ratio, hence higher stellar mass, than those with a more recent formation epoch due to the short-lived nature of massive bright stars. The masses derived from the BC and BaSTI models are found to be systematically higher at all metallicities and ages compared to those derived from MAR by $\simeq 0.1$ dex. These results are reasonably consistent with those found by other authors \cite{mar2006, wel2006,ret2006}. The choice of stellar population model can therefore have a significant affect on the derived BCG masses, meaning the significance of our result can vary between: $0.32\%-1.11\%$ for BC; $1.0\%- 9.8\%$ for MAR and $0.68\% - 2.5\%$ for BaSTI (Supplementary Table~1). If we restrict the metallicity to $Z \geq 1.0\, Z_{\odot}$  the significances remain greater than: $0.82\%$, $4.42\%$ and $2.53\%$ respectively for $z_f= 2.0$. These significance limits improve further if we let metallicity vary over its full range and instead restrict the formation epoch of these systems to $z_f\geq3$, with values of $0.58\%$, $2.25\%$ and $1.48\%$ respectively.

Turning to the observations, estimates of metallicities for BCGs in the literature are consistent with solar or super-solar abundancies \cite{humphrey2006, trager2008}. Furthermore, in addition to the $J-K_{\rm s}$ colour in Fig.~2 of the main text, evidence that the stars in BCGs form early is supported  by other investigations using the spectrophotometric properties of passively evolving BCGs and red galaxies at the centres of clusters. For example, a best fit model for the J0223 BCG from spectral template fitting using SSP models gives $z_f\simeq3$ (ref.~11); while a study of the CMR in the Lynx cluster \cite{Mei06} (J0849) at  $z=1.26$ implies a mean stellar age of 3.2 Gyr, corresponding to  $z_f>3.7$, with the stellar content of the bright elliptical galaxies in place and formed by $z\sim3$. Similar conclusions are reached for the early-type galaxies in J1252 from two studies of the colour-magnitude relation, which both suggest $z_f=2.7-3.6$ with subsequent passive evolution \cite{Blak03, lidman2004, demarco07}. We conclude that since observations suggest $Z \geq1.0 \, Z_{\odot}$ and $z_f\geq 3.0$, our central result of  negligible stellar mass evolution is not seriously compromised by the age and metallicity caveats and remains significant for the most plausible stellar evolution histories of these galaxies.

In order to further investigate systematic effects, some authors \cite{wel2006,ret2006} have attempted to calibrate stellar mass from the SSP models  with independent estimates of gravitational mass from velocity dispersion measurements. This has uncovered potentially serious systematic offsets ($0.3-0.4$ dex) between the two mass estimators for galaxies at $z\sim1$.
Such offsets are currently difficult to interpret as there is evidence that they may equally well be produced by biases in dynamical mass measurements or evolution in the dynamical structure of galaxies, as by biases in the photometric stellar mass determinations \cite{ret2006}. However, we note that the hybrid model shown in Fig.~2 of the main text (equivalent to an exponentially decaying model with $\tau=0.93$ Gyr) is very close to the $\tau =0.97$ Gyr model used in one of these comparisons \cite{wel2006} which shows negligible offset ($0.03\pm0.06$ dex) between the stellar and kinematic mass estimates.

\newpage
\begin{tabular}{p{45mm}p{45mm}p{45mm}}
a&                      b          &       c
\end{tabular}
\vspace{-20pt}\begin{figure}[h] 
   \centering
\includegraphics[width=0.33\textwidth]{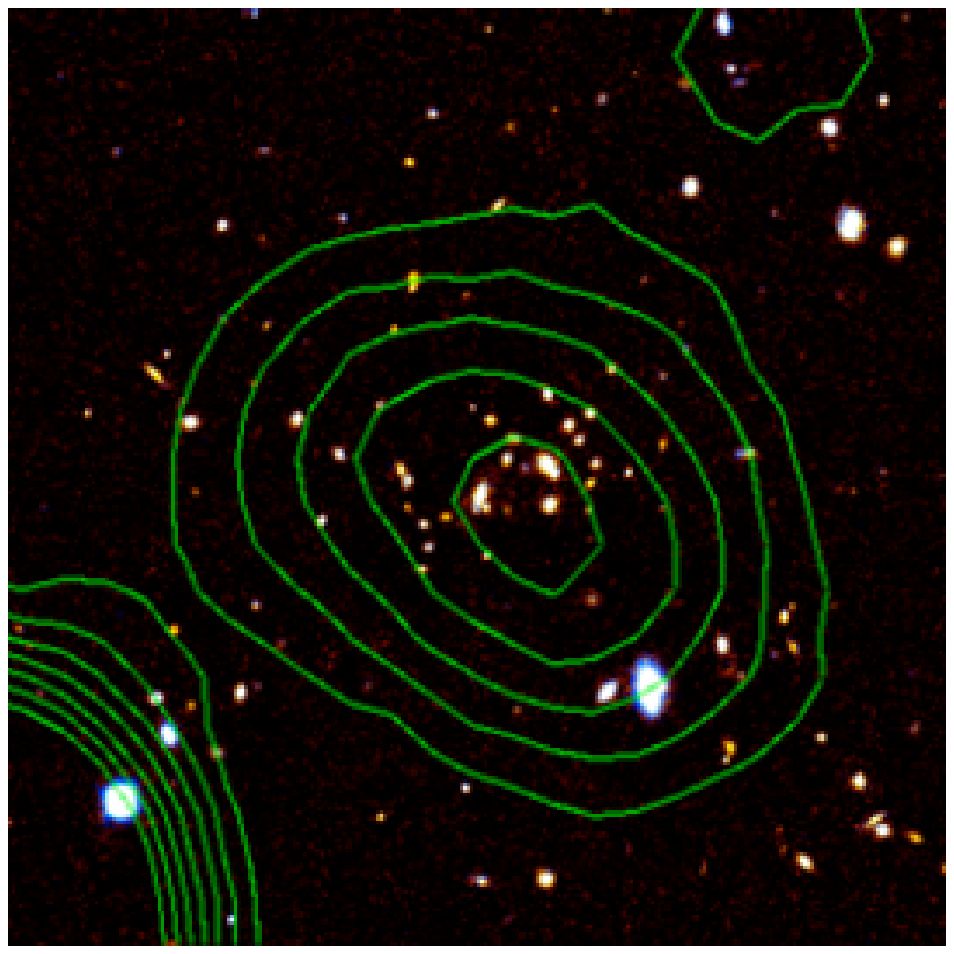}\includegraphics[width=0.33\textwidth]{2235cut.ps}\includegraphics[width=0.33\textwidth]{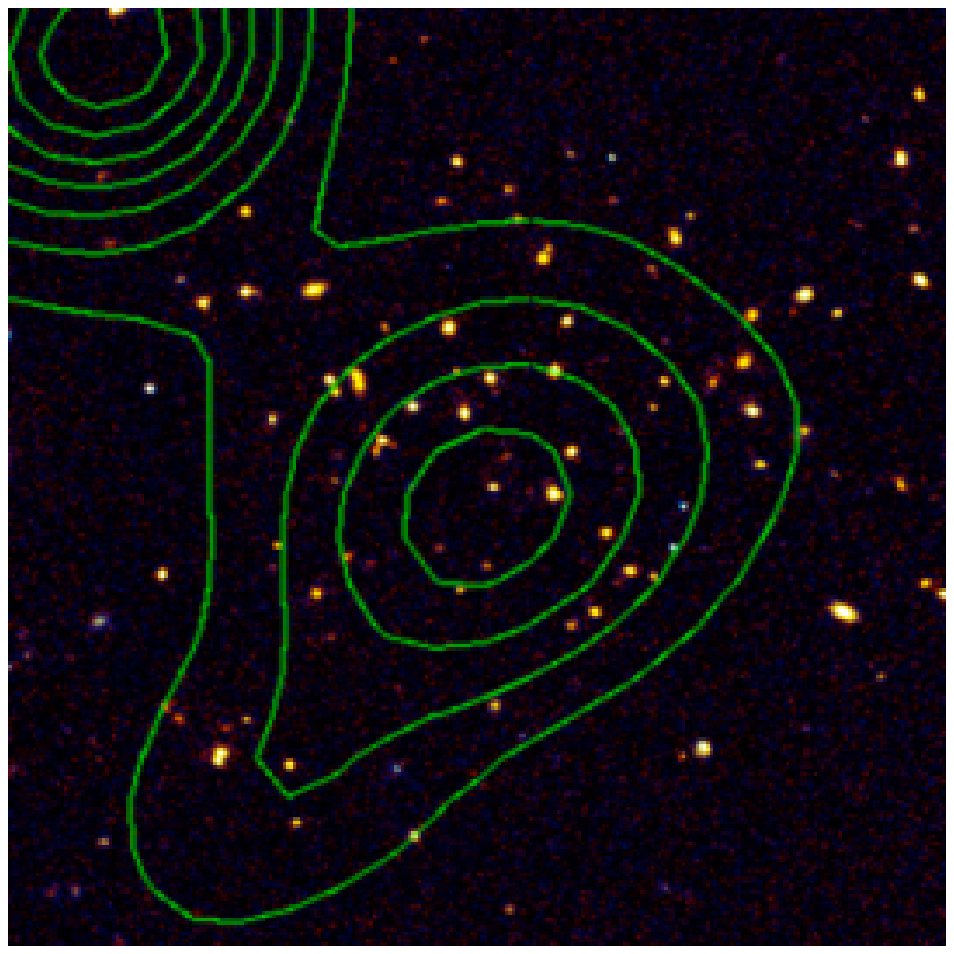}\\

{\vspace{40pt}{\bf Supplementary Figure~1:} Infrared images of the three high redshift clusters observed with the Subaru telescope at the summit of Mauna Kea. Data were taken in the $J$ and $K_{s}$ wavebands using the MOIRCS (Multi-Object Infrared Camera and Spectrograph) instrument at the Cassegrain focus. The images show the $1.5' \times 1.5'$ regions around the clusters. {\bf a,} XLSS J022303-043622. {\bf b,} XMMXCS J2235.3-2557. {\bf c,} XMMXCS J2215.9-1738. The X-ray contours overlaid in green are taken from the XMM-Newton XCS pipeline, smoothed with a Gaussian kernel and normalised to the peak of the cluster emission. At these redshifts $1.5'$ corresponds approximately to 0.75~Mpc.}
\end{figure}

\pagebreak
\begin{figure}
   \centering
\includegraphics[bb=275 0 300 300]{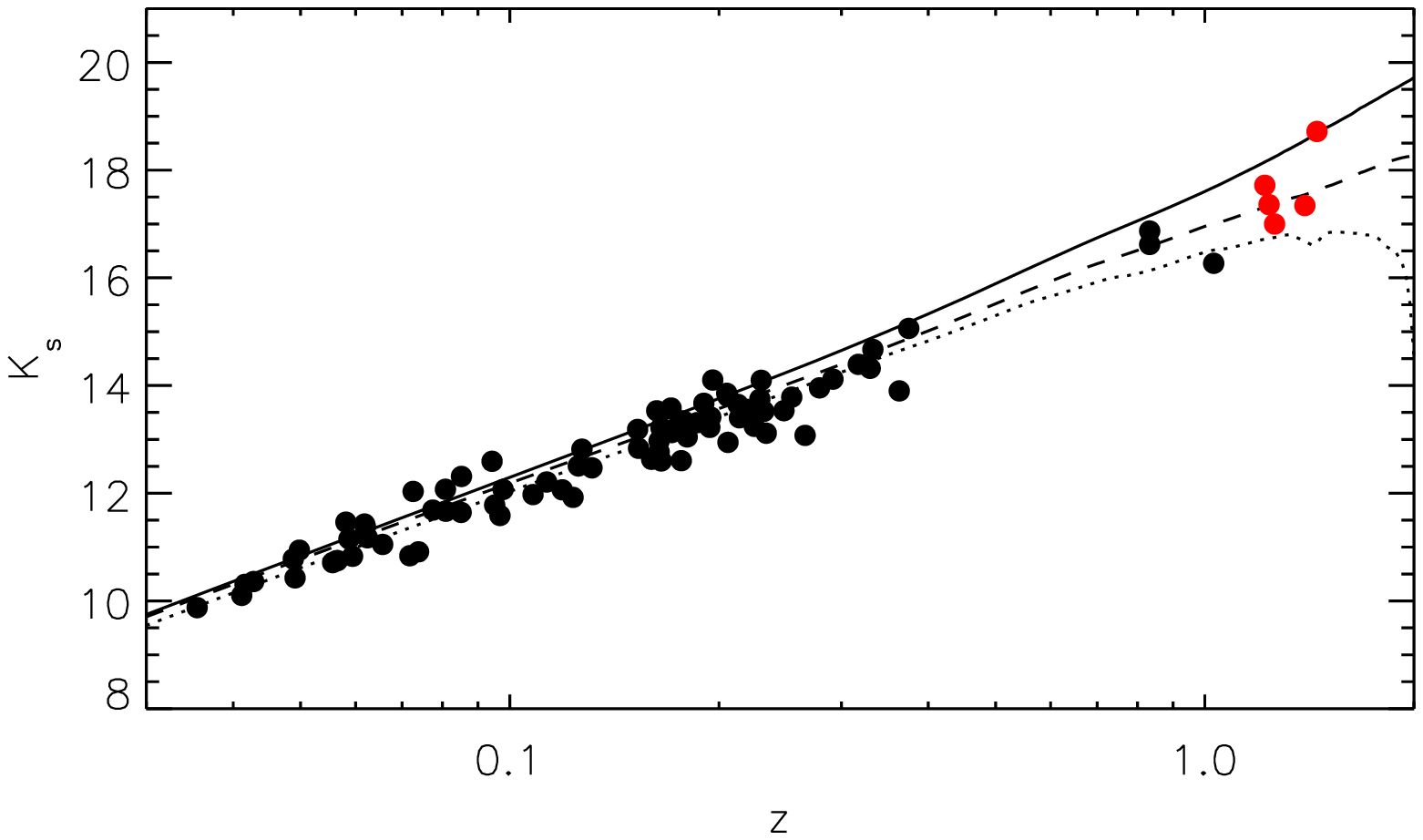} \\
{{\bf Supplementary Figure~2:} The BCG $K_{s}$ band Hubble diagram with the five high redshift BCGs shown in red. The photometric errors are $\simeq$0.01 mag as shown in Table~1 of the main text. The lines represent predictions of the observed $K_{s}$ magnitude based purely on stellar population models with: no evolution (solid), passive evolution with $z_f=$2 (dotted), the hybrid passive evolution model with $50\%$ of stars formed by $z=5$ and $80\%$ formed by $z=3$ (dashed), equivalent to an exponential decay with $\tau=0.93$. The black points indicate the comparison sample of 81 low redshift BCGs with masses in the same range as our high redshift sample.}
\end{figure}

\pagebreak
\begin{figure} 
   \centering
  \includegraphics[bb=200 0 300 300]{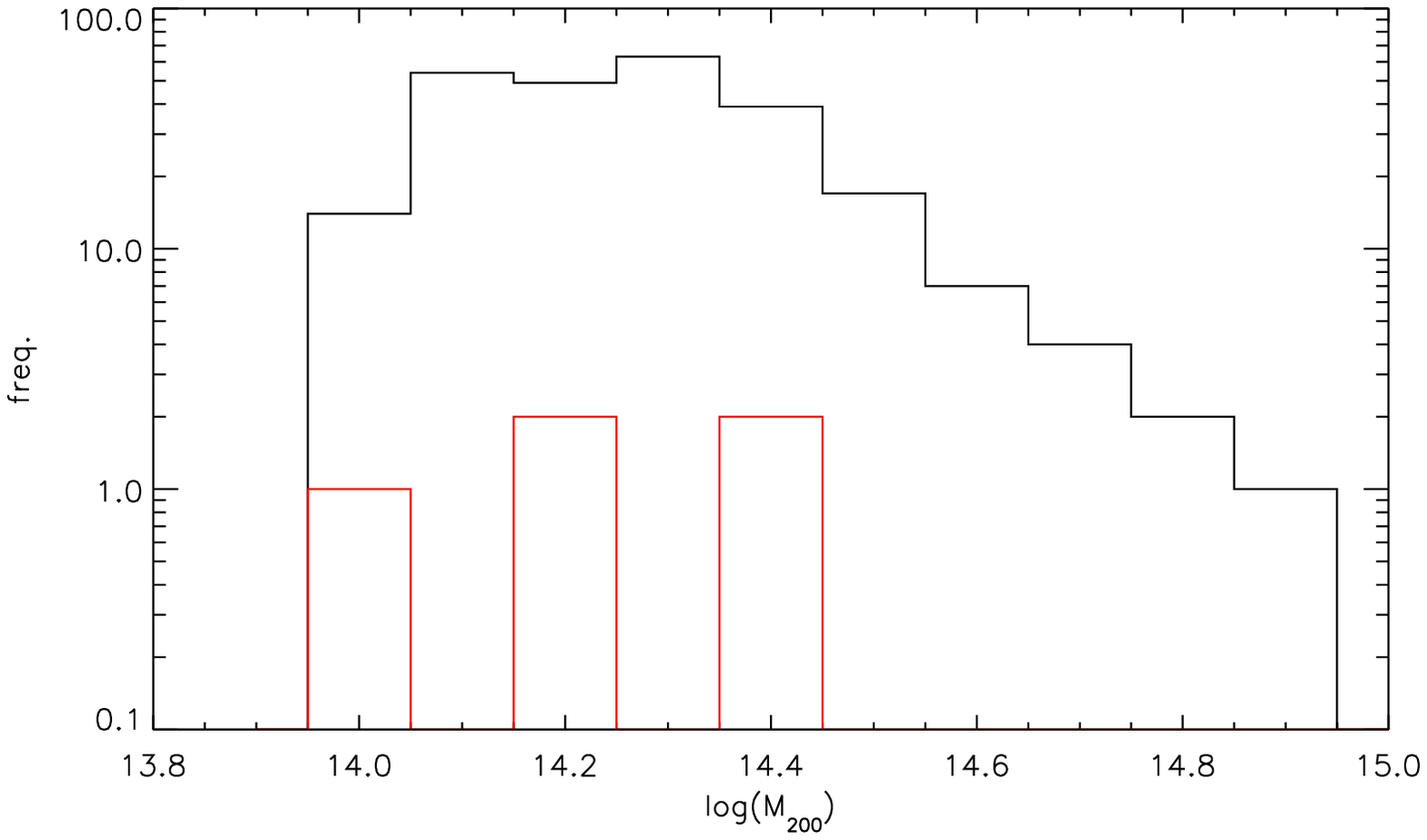} \\
  {\vspace{20pt}{\bf Supplementary Figure~3:}  A histogram of the 250 halo $M_{200}$ values from the semi-analytic model$\,^4$ at $z_f=1.08$ and $z_f=1.5$ (black) and the corresponding values for our observed sample (red). The average mass of the simulated and real cluster samples is $2.3\, (\pm 1.1) \times 10^{14}\, M_{\odot}$ (s.d.) and $2.1\,(\pm 0.8) \times10^{14}\,M_{\odot}$ (s.d.) respectively. This demonstrates that our cluster masses are well matched to those in the models allowing for direct comparison of the BCGs.}
\end{figure}

\begin{figure} 
   \centering
  \vspace{-50pt}
 \includegraphics[bb=0 10 200 180]{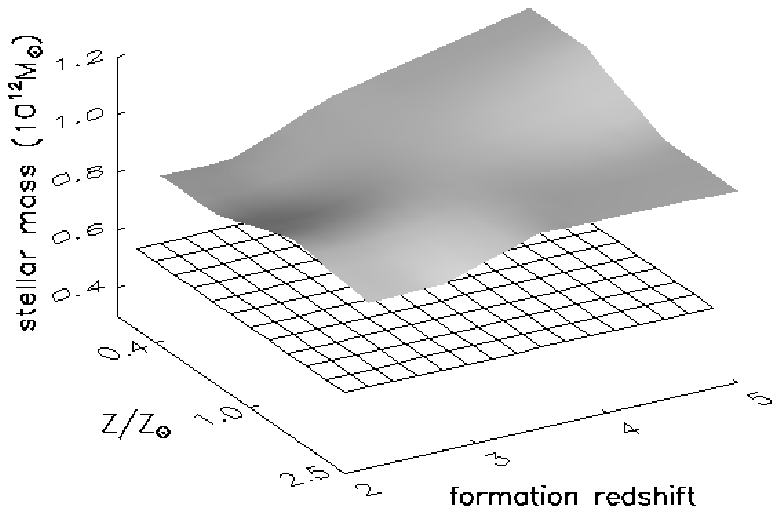} \\
  \includegraphics[bb=0 10 200 180]{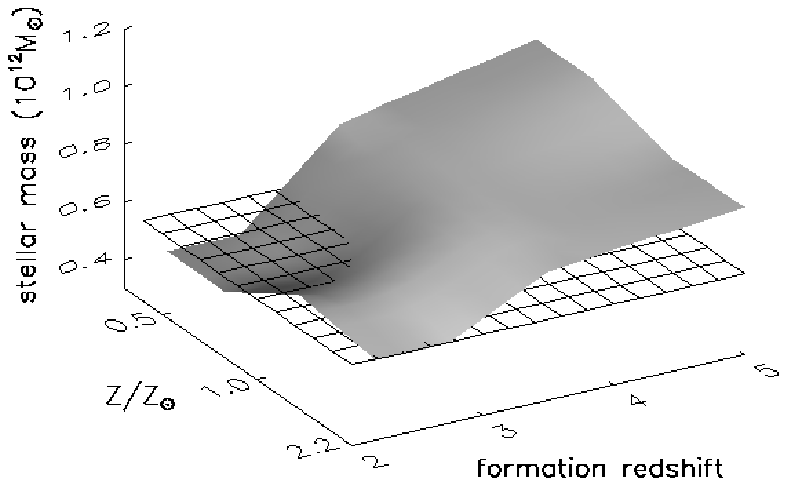}\\
  \includegraphics[bb=0 10 200 180]{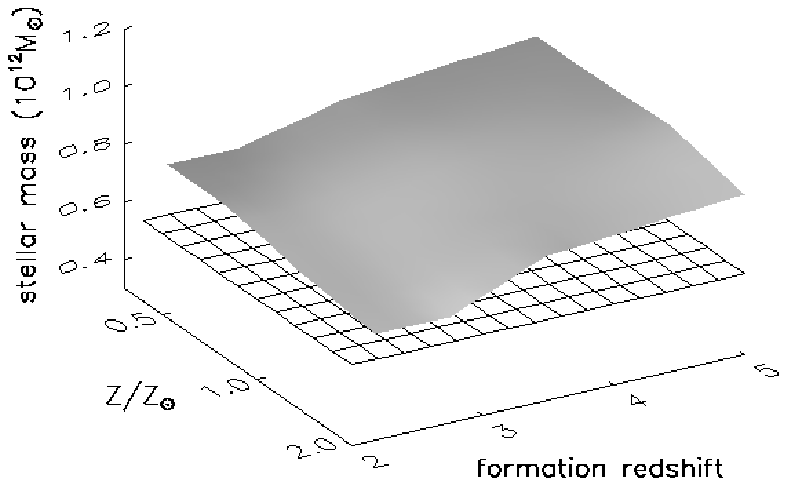}\\

{{\bf Supplementary Figure~4:} The three plots display the stellar mass surface obtained when the metallicity and formation redshift are varied over the ranges $Z=0.4Z_{\odot}-2.5Z_{\odot}$ and $z_f=2,2.5,3,5$ respectively. The models considered are BC, MAR and BaSTI (upper, middle, lower) as described in the Supplementary Information text. The grey shaded surface is the mean mass of our BCG sample while the black mesh surface represents a constant stellar mass of $5.55\times10^{11}M_{\odot}$ corresponding to a significance for a one-tailed $P$ of 0.05 ($\rm{d.f.}=4$) in the offset from the predicted BCG average mass. The masses derived from the BC and BaSTI models are found to be well in excess of this limit for all combinations of metallicity and age considered here. For the MAR model there is shown to be a steep drop in mass-to-light ratio for young stellar populations meaning this significance (0.05) is not achieved if both the metallicity $Z < 1Z_{\odot}$ and  stellar population formation epoch $z_f < 2.5$ (see Supplementary Table 1).}
\end{figure}

\begin{table*}\footnotesize
\begin{center}
\label{tab:samplecm}
\begin{tabular}{rcccc}
\hline
&$z_{f}=5$&$z_{f}=3$&$z_{f}=2.5$&$z_{f}=2$\\
\hline
\noalign{\medskip}
\noalign{\smallskip}
\multicolumn{2}{l}{BC \hfil}\\
\noalign{\smallskip}
1.0Z$_{\odot}$&0.97(0.54)&0.96(0.58)&0.93(0.63)&0.91(0.71)\\
0.4Z$_{\odot}$&1.10(0.32)&0.96(0.58)&0.93(0.63)&0.82(1.11)\\
2.5Z$_{\odot}$&0.97(0.54)&0.98(0.52)&0.94(0.62)&0.88(0.82)\\
\noalign{\medskip}
\multicolumn{2}{l}{MAR \hfil}\\
\noalign{\smallskip}
1.0Z$_{\odot}$&0.75(1.61)&0.70(2.13)&0.63(3.23)&0.62(3.43)\\
0.5Z$_{\odot}$&0.84(1.00)&0.70(2.13)&0.53(6.12)&0.46(9.82)\\
2.2Z$_{\odot}$&0.76(1.52)&0.69(2.25)&0.59(4.14)&0.58(4.42)\\
\noalign{\medskip}
\multicolumn{2}{l}{BaSTI \hfil}\\
\noalign{\smallskip}
1.0Z$_{\odot}$&0.92(0.68)&0.80(1.48)&0.77(1.44)&0.70(2.13)\\
0.5Z$_{\odot}$&0.91(0.71)&0.84(1.00)&0.82(1.11)&0.76(1.52)\\
2.0Z$_{\odot}$&0.86(0.91)&0.80(1.23)&0.76(1.52)&0.67(2.53)\\
\hline
\end{tabular}
\end{center}
{{\bf Supplementary Table~1:}  The matrix of BCG stellar mass values (in units $10^{12}M_{\odot}$) for our  BCG sample assuming $z=1.3$ derived for a set of formation redshifts ($z_{f}$) and metallicities for the three stellar population codes BC, MAR and BaSTI, as described in the Supplementary Information text. The numbers in brackets are the significance percentages ($\rm{d.f.}=4$) of the masses compared to the average ($0.192\,(\pm 0.038)\times10^{12}M_{\odot}$) of the 250 simulated BCGs. The stellar code used in the semi-analytic model$^{4}$ is based on BC and almost identical to that shown in Fig.~2 of the main text; this gives a stellar mass of $0.899\,(\pm0.082) \times10^{12}M_{\odot}$. Comparing the $J-K_{\rm s}$ colours of these models with the data we find that for $z_{f}\leq2$ with either solar or sub-solar metallicity the BC models are too blue at the $3\sigma$ level compared to the average colour of our sample ($J-K_{\rm s}=1.84 \pm 0.05$, s.d.). Otherwise the colours are reasonably consistent providing no additional significant constraint.}
\end{table*}

\vspace{1000pt}{\bf\Large References}
\bibliography{xcs}

\end{document}